\documentclass[aps,prb,twocolumn,showpacs,amsfonts,floatfix]{revtex4-2}
\usepackage{epsfig}
\usepackage[dvipsnames]{xcolor}
\usepackage{graphics}
\usepackage{graphicx}
\usepackage{amsmath,ulem}
\usepackage{amssymb,multirow}
\usepackage{bm}
\usepackage{color}
\usepackage[utf8]{inputenc}

\begin{document}

\title{Inter-quintuple layer coupling and topological phase transitions in the chalcogenide topological insulators}

\author{K. Shirali}

\author{W. A. Shelton}

\author{I. Vekhter}
\affiliation{Department of Physics and Astronomy, Louisiana State University, Baton Rouge, LA 70803-4001}

\date{\today}

\begin{abstract}
Driving quantum phase transitions in the 3D topological insulators offers pathways to tuning the topological states and their properties. We use DFT-based calculations to systematically investigate topological phase transitions in Bi$_2$Se$_3$, Sb$_2$Se$_3$, Bi$_2$Te$_3$ and Sb$_2$Te$_3$ by varying the $c/a$ ratio of lattice constants. This ensures no net hydrostatic pressure under anisotropic stress and strain and allows a clear identification of the physics leading to the transition. As a function of $c/a$, all of these materials exhibit structural and electronic stability of the quintuple layers (QLs), and quasi-linear behavior of both the inter-quintuple layer distance and the energy gap near the topological transition. Our results show that the transition is predominantly controlled by the inter-QL physics, namely by competing Coulomb and van der Waals interactions between the outer atomic sheets in neighboring quintuple layers. We discuss the implications of our results for topological tuning by alloying.
\end{abstract}


\maketitle

\section{Introduction}
Topological insulators~\cite{Hasan2010,Hasan2011,Qi2011,Ando2013} (TIs) possess conducting linearly-dispersing surface states which are protected by time-reversal symmetry. In the simplest cases the two atomic-derived energy levels closest to the chemical potential (after accounting for the crystal-field splitting and hybridization) have opposite parity, and are sufficiently close in energy so that Spin-Orbit Coupling (SOC) can flip the energies of those localized levels, and induce inversion between the valence and conduction bands derived from these states in the Brillouin Zone (often close to the $\Gamma$-point)~\cite{bernevig2006quantum,Zhang2009,Liu2010}. The SOC thus competes with the initial level splitting: if its magnitude is sufficient to generate an inverted band structure, there exists a non-zero bulk topological invariant that guarantees the existence of protected surface states. These states survive as long as the band gap persists, and a transition to a topologically trivial state requires closing of the energy gap.

Pressure and strain have emerged as promising avenues towards controlling topological properties~\cite{Liu2014,Zeljkovic2015,Flototto2018,Yang2018}. These may result from alloying or doping by elements of different atomic size that gives rise to chemical pressure\cite{Zhang2011_al,Kong2011,Ko2013,Post2015,Tung2016,Nam2019}. They also appear in heterostructures due to lattice mismatch. Both strain and pressure leave the (atomic) SOC strength mostly unchanged, but modify the hybridization of atomic levels and crystal-field strength. As a result, they may induce a transition between a topological and a trivial state. The physics of this transition is the focus of our work.

We use the ratio of the out-of-plane to in-plane lattice constants in layered compounds, $c/a$, as a tuning parameter that drives the topological transition, and study it using ab initio Density Functional Theory calculations. The advantage of such an approach is that the system is under no net hydrostatic pressure, albeit simultaneously experiences uni- or bi-axial pressure and strain. This allows us to focus on the physics driving the transition.



We apply this methodology to the family of layered tetradymite materials X$_2$Y$_3$ (X=Bi/Sb, Y=Se/Te). Among those, Bi$_2$Se$_3$, Sb$_2$Te$_3$, and Bi$_2$Te$_3$ have been confirmed via experiment~\cite{Xia2009,Ando2013,Chen2009,Hsieh2009a,Hsieh2009b,Zhang2009} to be topological insulators. The fourth member of this family, Sb$_2$Se$_3$, has not been grown in the same rhombohedral crystal structure, but was initially predicted to be a trivial insulator~\cite{Zhang2009,Liu2011,Li2014,LiuJ2014}. More recent theoretical studies~\cite{Aramberri2017,Cao2018} predicted the ground-state of rhombohedral Sb$_2$Se$_3$ to be topological. Our results confirm this, and we therefore focus on the common features of the topological transitions in all four members of the family.

The application of pressure was predicted to induce topological phase transitions in the 3D TIs 
in a number of first-principles~\cite{Liu2011,Young2011,Bera2013,Li2014,Aramberri2017,Saha2021,Reid2021} and analytical~\cite{Lee2015,Brems2018} studies. However, a systematic physical picture has not emerged, in part because of strongly anisotropic and implementation-dependent~\cite{Kim2018}, response of the structural parameters to the pressure and strain. This motivated our choice of the $c/a$ ratio as the tuning parameter that eliminates the hydrostatic pressure effects.

For each of the Bi/Sb Te/Se compounds we find a critical value of $c/a$ where the band gap closes, and beyond that value the band inversion disappears so that the material becomes topologically trivial. In the vicinity of the transition all four materials exhibit similar behavior. In particular, the phase transition is due to the increased level splitting which prevents SOC from inducing the band inversion, and we find that the underlying physics is dictated by the Coulomb and van der Waals interactions between the quintuple layers, which are the basic structural building blocks of these materials. The quintuple layers themselves remain essentially unchanged across the transition, with minimal variation in the overall thickness and bond lengths. Consequently, the hybridization does not vary significantly, in contrast to the often used simplified picture. Instead, it is the inter-quintuple layer interaction that changes the energies of the outer (Se/Te -derived) states compared to their inner (Bi/Sb-derived) counterparts. We discuss the universal features of this inter-quintuple layer interaction and their implication for transitions in other stoichiometric and alloyed topological insulators.

The rest of the paper is organized as follows: in Section II, we describe the family of 3D TIs and provide the computational details and methods used to carry out the analysis. In Section III, we discuss the topological phase transition in Bi$_2$Se$_3$ and in Section IV, the topological phase transitions in Sb$_2$Se$_3$, Bi$_2$Te$_3$ and Sb$_2$Te$_3$. Section V summarizes our findings, and discusses their implications for other topological systems.

\begin{figure}
    \centering
    \includegraphics[trim={0 2.75cm 2cm 9cm},clip,width=\columnwidth]{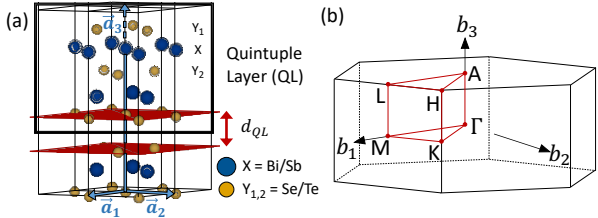}
    \caption{Crystal structure of X$_2$Y$_3$ (X=Bi/Sb, Y=Te/Se) with (a) hexagonal unit cell that consists of 3 quintuple layers (QLs). We show the in-plane lattice vectors $\vec{\bm a}_{1,2}$. The lattice vector $\vec{\bm a}_3$ in the direction normal to the QLs is not to scale. (b) Hexagonal Brillouin zone.}
    \label{fig:bulk}
\end{figure}

\section{Technical and Computational Details}
Bi$_2$Se$_3$, Sb$_2$Se$_3$, Bi$_2$Te$_3$ and Sb$_2$Te$_3$ are rhombohedral compounds, crystallographic group $R\overline{3}m$ $(166)$, point group $D_{3d}$. As shown in Fig.~\ref{fig:bulk}, they consist of stacked structural units of five atomic layers $Y1-X-Y2-X-Y1$ ($X$ = Bi/Sb, $Y_{1,2}$ = Te/Se) called quintuple layers (QLs). The atomic layers follow an $...ABCABC...$ stacking in the $c$-axis direction. The pairs of atoms $Y1$ are related to each other through inversion symmetry, as are the $X$ atoms. Each QL has a `closed-shell' nature, leading to stronger bonding within QLs that is believed to be mainly covalent, and a weaker bonding between QLs. While van der Waals interactions contribute significantly to the inter-QL coupling, the electronic bandwidths in-plane and in the direction normal to the layers are comparable, so that these are structurally and electronically three-dimensional materials, in contrast to, for example, transition metal dichalcogenides.

Calculations were carried out using the Vienna Ab initio Simulation Package~\cite{Kresse:1993,Kresse:1994,Kresse:1996a,Kresse:1996b} (VASP), version 5.4.4, and \textsc{Quantum ESPRESSO} v.6.7~\cite{QE-2009,QE-2017,doi:10.1063/5.0005082}. 

Crystallographic information for Bi$_2$Se$_3$~\cite{Nakajima1963} and Bi$_2$Te$_3$~\cite{Nakajima1963}, as well as for Sb$_2$Te$_3$~\cite{Anderson1974} is taken from experimental data retrieved from Crystallography Open Database~\cite{Merkys2016, Merkys2015,Merkys2012,Grazulis2009,Downs2003}. In our calculations we use the Bi$_2$Se$_3$ crystal structure of Nakajima~\cite{Nakajima1963}, rather than the earlier data of Refs.~\onlinecite{semiletov1955electron,wyckoff1964structures}. A comparison of the two can be found in Ref.~\onlinecite{Luo2012}. For Sb$_2$Se$_3$, we use a starting guess for the lattice parameters based on the values reported in Ref.~\onlinecite{Li2014}.

For VASP calculations we used Projector Augmented Wave (PAW) potentials~\cite{Blochl:1994,Joubert:1999} for Bi ($6s^{2} 6p^{3}$), Sb ($5s^{2} 5p^{3}$), Se ($4s^{2} 4p^{4}$) and Te ($5s^{2} 5p^{4}$), and a plane-wave basis. We also used PBE-GGA~\cite{PerdewBurkeErnzerhof:1996,PerdewBurkeErnzerhof:1997} for the exchange-correlation functional, and van der Waals interactions were included using the DFT--D2~\cite{Grimme2006} method, following the framework in Ref.~\onlinecite{Shirali2020}. We confirmed that inclusion of Bi $d$ electrons does not change our results for Bi$_2$Se$_3$. We used a $\Gamma$-centered k-point grid of 11x11x11 k-points and an energy cut-off of 450 eV
for the plane wave basis. 

We performed relativistic calculations which include spin-orbit coupling (SOC), with the convergence threshold for energy of $10^{-5}$ eV. In our calculations we used 15 atom hexagonal unit cells for Bi$_2$Se$_3$, Sb$_2$Se$_3$, Bi$_2$Te$_3$ and Sb$_2$Te$_3$.

To confirm the validity of our results we also ran a calculation using \textsc{Quantum ESPRESSO}, using the pseudopotentials Bi.rel-pbe-dn-kjpaw\_psl.1.0.0.UPF and Se.rel-pbe-n-kjpaw\_psl.1.0.0.UPF from \url{https://www.quantum-espresso.org}, k-grid of 15x15x15 k-points, and energy cut-offs of 653 eV and 6530 eV for the wave-functions and charge densities respectively. We found no substantial differences with the results obtained in VASP.


To calculate the ground-state lattice parameters of each material, we run full structural optimization in which the cell shape, volume, and atomic coordinates are allowed to relax. To investigate the dependence of the structural and electronic properties on $c/a$, we fix this ratio and carry out structural optimization by running alternate atom and volume relaxations (ISIF=2,7), while not allowing the cell shape to change.

Band structures are plotted with data processed using vaspkit~\cite{Wang2021}, while the band structure projections are plotted using PyProcar~\cite{HERATH2020107080}. Since we plot the electron bands along a chosen path in the Brillouin Zone, band gaps were also estimated from the density of states.

For Sb$_2$Se$_3$ slab calculations, the surfaces are modelled using a 12 QL-thick ($\sim$ 120 \AA) slab. With vdW interaction included we find that using a vacuum thickness of $\sim$ 100 {\AA} is sufficient to avoid interaction of the slab with periodic images of itself. Atoms in the three outermost QLs of the slab are allowed to relax in all directions without restriction, while atoms in the ``bulk'' part of the slab are held fixed.

\section{Topological phase transition in B\lowercase{i}$_2$S\lowercase{e}$_3$}

\begin{figure}
    \centering
    \includegraphics[width=\columnwidth,trim={1cm 0.5cm 1.5cm 1.0cm},clip]{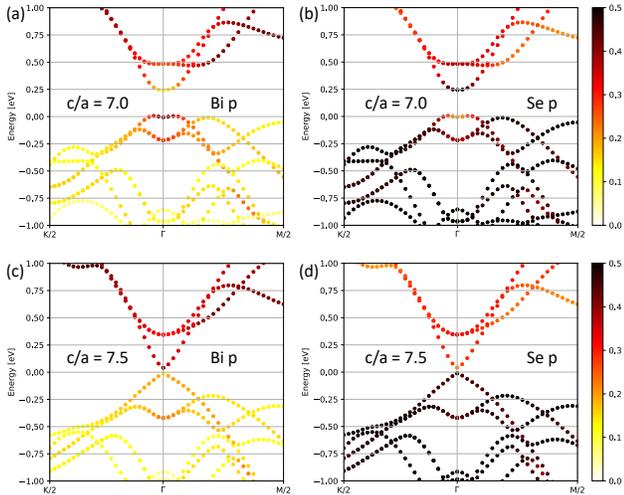}
    \caption{Bulk band structure of Bi$_2$Se$_3$ projected onto Bi $p$ and Se $p$ for different $c/a$ values. (a), (b) Band structure projections onto Bi $6p$ and Se $4p$ for $c/a$ = 7.0 (topological insulating state). (c), (d) Corresponding projections for $c/a$ = 7.5 (trivial insulating state). The bands below the chemical potential in (a), (c) have very small weights of Bi $p$ away from $\Gamma$, and hence have a very light shade.}
    \label{fig:bulk_projections_Bi2Se3}
\end{figure}

In the absence of external strain, Bi$_2$Se$_3$ is a topological insulator. Minimization of the energy gives the ground-state at $c/a$ = 6.91~\cite{SI}. In the ground-state, the system is topological, and its bands are inverted: the valence (conduction) band is dominated by the Bi $6p_z$ (Se $4p_z$) orbitals at $\Gamma$, but the orbital contribution to the bands switches away from the center of the Brillouin Zone. This inversion persists as we increase the $c/a$ ratio.  Figure~\ref{fig:bulk_projections_Bi2Se3}(a), (b) shows the bulk bands projected onto the Bi $p$ and Se $p$ orbitals for $c/a$ = 7.0, with the weights depicted according to the color scale on the right, and demonstrates this inversion. As expected on general grounds, the band inversion is always accompanied  by  the emergence of  linearly dispersing surface states, see Ref.~\onlinecite{Shirali2020} and Sec.~\ref{sec:Sb2Se3}, and therefore we use it as a test for the topological electronic structure.

We confirmed that if we do not include relativistic SOC in our first-principles calculations, the bands are not inverted, and that the conduction band is composed of  Bi $p_z$ orbitals, whereas the valence band is composed of  Se $p_z$ orbitals. 

When the $c/a$ ratio is increased, upon relaxing the unit cell, the in-plane lattice constant $a$ decreases whereas the lattice constant $c$ increases. The detailed results are given in the Supplementary Material~\cite{SI}. Between $c/a$ = 7.0 and $c/a$ = 7.5 the value of $a$ ($c$) decreases (increases) by 2\% (6\%), see~\cite{SI}. Within the QL, the Bi-Se1/Bi-Se2 bond lengths change by less than $0.4\%$ 
over the same range of $c/a$ values.
As the out-of-plane lattice constant $c$ increases, the most significant change happens in the inter-QL distance $d_{QL}$ which grows 
by 13\%, whereas the QL-thickness $h_{QL}$ only increases by 3\% between the same $c/a$ values. Thus, the change in the inter-QL distance is about two orders of magnitude larger than the magnitude of changes within the QL.

\begin{figure}
    \centering
    \includegraphics[width=\columnwidth,trim={2.5cm 0.5cm 2.5cm 1cm},clip]{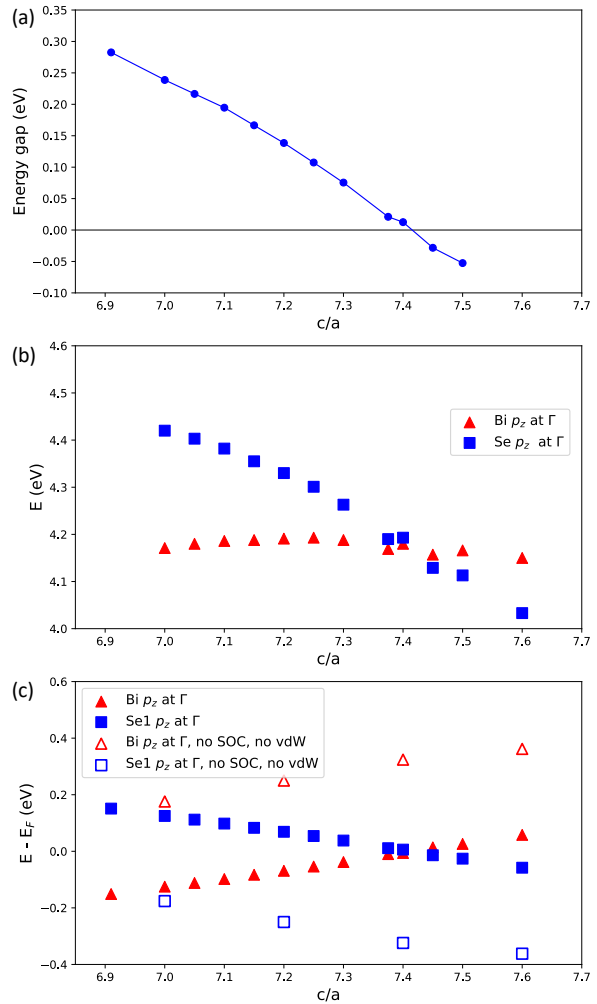}
    \caption{Evolution of parameters with $c/a$ for the topological phase transition in Bi$_2$Se$_3$. (a) Variation of the band gap; (b) change in the Bi $p_z$ and Se $p_z$ states' energies at $\Gamma$; (c) change of these energies relative to the chemical potential. Red (blue) filled symbols show the energies of the Bi $p_z$ (Se $p_z$) levels for calculations including SOC and vdW interactions, whereas red (blue) unfilled symbols in panel (c) show the energies of the Bi $p_z$ (Se $p_z$) levels for calculations in which neither SOC nor vdW interactions are included.}
    \label{fig:transition_Bi2Se3}
\end{figure}

As the $c/a$ ratio is increased, the band gap at the $\Gamma$ point, $E_g$, decreases, and closes at a critical value. Beyond this point, upon further increase of $c/a$, the gap reopens, but the energy bands do not show inversion -- the conduction band is Bi $6p_z$-derived, and the valence band is mostly Se $4p_z$-derived throughout the Brillouin Zone,
as shown in Fig.~\ref{fig:bulk_projections_Bi2Se3}(c), (d). Thus, the system undergoes a transition from a topological state to a trivial state as $c/a$ is increased. In Fig.~\ref{fig:transition_Bi2Se3}(a), we plot the evolution of the bulk band gap $E_g$ with $c/a$, and assign positive (negative)  values to the case of inverted (non-inverted) bands. The gap varies almost linearly within the range we consider, and the critical $c/a$ at which the transition occurs is approximately $7.42$.


In Fig.~\ref{fig:transition_Bi2Se3}(b), we show the evolution of the absolute (not relative to the chemical potential) energies of the Bi $p_z$- and Se $p_z$-derived states at $\Gamma$ as a function of the $c/a$ ratio. 
We find that the energies of the Bi $p_z$ levels at $\Gamma$ are almost unchanged as $c/a$ is increased (slope 0.03 $\pm$ 0.02 eV)
In contrast, the energy of Se $p_z$ levels varies considerably, and is nearly linear in $c/a$ with the slope -0.65 $\pm$ 0.03 eV.  Since the slope of the Bi $p_z$-level is very close to zero, and the separation between the Bi and Se $p_z$-levels is almost equal to the band gap (though the `camelback' structure near the valence band maximum, see Fig.~\ref{fig:bulk_projections_Bi2Se3}, reduces the band gap slightly compared to this separation), the slope of the Se $p_z$ levels in Fig.~\ref{fig:transition_Bi2Se3}(b) is very close to that derived from the evolution of the band gap in Fig.~\ref{fig:transition_Bi2Se3}(a), which is -0.58 $\pm$ 0.01 eV.

To investigate the roles of the relativistic spin-orbit coupling and van der Waals interactions, at each value of $c/a$, we turn off these two effects in our calculations, optimize the lattice parameters to stay in the ground state, and monitor the variations in the Bi $p_z$ and Se $p_z$ levels at $\Gamma$. Inclusion of van der Waals interactions changes the lattice constants and other structural parameters~\cite{Luo2012,Cheng2014,Shirali2020} and the total energy for each $c/a$. Therefore we find that the chemical potential shift between the calculations with and without vdW and SOC is also a function of the $c/a$ ratio. To facilitate a comparison between the two cases, we now reference the energies of both Bi and Se states to the respective chemical potentials, set in the middle of the gap at $T=0$, and plot both in Fig.~\ref{fig:transition_Bi2Se3}(c). 

Importantly, we find that the change in the energy of these states near the $\Gamma$-point 
depends only weakly on the SOC and vdW interactions. In Fig.~\ref{fig:transition_Bi2Se3}(c), the empty (filled) symbols show the Bi (triangles) and Se (squares) $p_z$ level energies at $\Gamma$ without (with) SOC and vdW. 
As expected, we observe the inversion: without SOC+vdW, the Bi (Se) state is above (below) the chemical potential. Inclusion of spin-orbit interaction inverts the levels and leads to a topological state. What is striking in Fig.~\ref{fig:transition_Bi2Se3}(c)  is that the evolution (the slope) 
is nearly identical for both cases. 
The slopes of the Se $p_z$ levels in Fig.~\ref{fig:transition_Bi2Se3}(c) are -0.306 $\pm$ 0.004 eV and -0.32 $\pm$ 0.03 eV with and without SOC+vdW respectively. The slopes of Bi levels are also essentially the same, but with opposite signs. We conclude that the \textsl{change in the energy of} the states closest to the chemical potential does not depend strongly on SOC+vdW corrections. 

In summary, as we tune Bi$_2$Se$_3$  towards the topological transition, the quintuple layers remain largely intact, while the interlayer distance changes significantly. The energy of the states dominated by the electrons in the outer (Se) atomic planes changes much more than the energy of those originating in the inner (Bi) layers, suggesting that the interlayer interaction is the main driver of the transition. Finally, 
the variations of these energies with $c/a$ are not affected by the vdW and SOC corrections (even though the absolute energies are). We therefore conclude that 
the topological transition appears to be driven largely by the inter-QL Coulomb physics contained in the exchange-correlation functional. A natural question is whether 
these features are universal across the entire family of these compounds.



\section{Topological phase transitions in S\lowercase{b}$_2$S\lowercase{e}$_3$, B\lowercase{i}$_2$T\lowercase{e}$_3$ and S\lowercase{b}$_2$T\lowercase{e}$_3$}

\subsection{Predicted topological insulator Sb$_2$Se$_3$}
\label{sec:Sb2Se3}
Sb$_2$Se$_3$, in the rhombohedral structure identical to that of Bi$_2$Se$_3$, was initially thought to be a trivial insulator~\cite{Zhang2009,Liu2011,Bera2013,Li2014,LiuJ2014} in the ground state. 
These studies used the PBE-optimized structure in their calculations and did not include the van der Waals correction when obtaining this result (Ref.~\onlinecite{Liu2011} optimized the structure including the van der Waals correction, but did not use that structure for the rest of their calculations). The effects of applying uniaxial~\cite{Liu2011} and hydrostatic~\cite{Li2014} pressure on Sb$_2$Se$_3$ have been studied before. Ref~\onlinecite{Liu2011} found a topological phase under compressive longitudinal strain, when the lattice constant $c$ is decreased by more than 2\% from its equilibrium value. Ref.~\onlinecite{Li2014} found that under hydrostatic pressures close to and greater than 1.0 GPa, the material is driven into a topological phase.

More recent studies~\cite{Aramberri2017, Cao2018} included the van der Waals correction in the structural optimizations and predicted ground-state Sb$_2$Se$_3$ to be a topological insulator. They also studied the application of strain: Ref.~\onlinecite{Aramberri2017} included the vdW interactions in the form of the DFT+D2 method, and reported that under a tensile strain of 3\% along the c-axis, the material becomes a trivial insulator. Ref.~\onlinecite{Cao2018} ran structural optimizations using various vdW methods (optB86b-vdW, optB88-vdW, optPBE-vdW and DFT-TS), and found that under the application of a tensile strain, the material is driven into a trivial phase and loses the inverted band structure.

\begin{figure}
    \centering
    \includegraphics[width=\columnwidth]{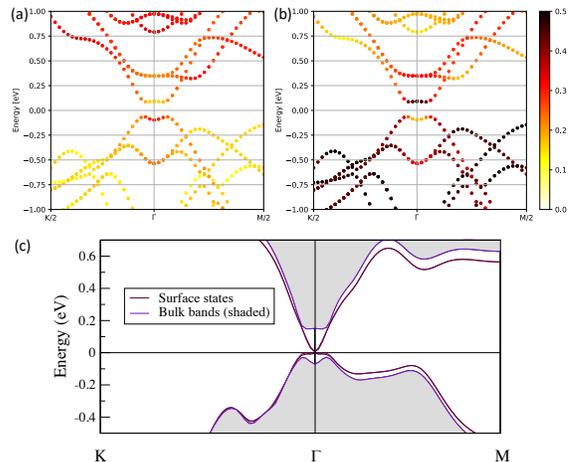}
    \caption{Bulk band structure projections and slab band structure for Sb$_2$Se$_3$. Band structure projections onto (a) Sb $5p$ and (b) Se $4p$ for $c/a$ = 7.137 (topological insulating state). The bands below the chemical potential in (a) away from $\Gamma$ have negligible contribution from Sb $p$, and can hence barely be seen. (c) Topological surface states for $c/a$ = 7.0 for a 12 QL-thick slab, with 100 {\AA} vacuum.}
    \label{fig:bulk_projections_and_slab_Sb2Se3}
\end{figure}

Our results confirm the conclusions of these studies. When we use PBE-GGA for structural optimization, we obtain the ground-state to be a trivial insulator, in agreement with Refs.~\onlinecite{Liu2011,Li2014}.
Upon including the vdW interactions in our calculations, we find that the ground-state of Sb$_2$Se$_3$ is a topological insulator (see~\cite{SI}), with a band gap of $E_g\approx 0.152$ eV. This can be seen in Fig.~\ref{fig:bulk_projections_and_slab_Sb2Se3}(a) and Fig.~\ref{fig:bulk_projections_and_slab_Sb2Se3}(b), which show the inversion of the bulk bands when projected onto the Sb $p$ and Se $p$ orbitals in the ground state,  $c/a$ = 7.137~\cite{SI}. 

To confirm the existence of the topological surface states for $c/a$ values in which Sb$_2$Se$_3$ exhibits inverted bands, we run slab calculations using both 5 QL-thick and 12 QL-thick slabs with vacuum thicknesses of 100 {\AA}. As an example, for $c/a$ = 7.0, the surface states in a 5 QL-thick slab are gapped, and their spatial profile reveals a significant overlap between states localized at opposite surfaces of the slab. Hybridization between the two surface states is almost eliminated for a 12 QL-thick slab, where we obtain the Dirac-like gapless topological surface states, shown in Fig.~\ref{fig:bulk_projections_and_slab_Sb2Se3}(c). The atoms which are allowed to relax only have movements of the order of m\AA  (milli-Angstrom), confirming that the slabs are stable and do not suffer from large forces. The longer decay length, $\sim$ 3 QLs, in Sb$_2$Se$_3$ compared to $\sim$ 1 QL 
in Bi$_2$Se$_3$ and 
Bi$_2$Te$_3$, is in agreement with Ref.~\onlinecite{Cao2018}, and is consistent with having a smaller bulk energy gap than in Bi-based materials.

When we run slab calculations using a 12 QL-thick slab using the optimized structure for $c/a$ = 7.549, a value where the material does not display band inversion in the bulk, as expected we find that there are no topological surface states. Thus, we confirm that the band inversion in our calculations is a reliable test for the topological nature of these materials. 

\subsection{Topological phase transitions in Sb$_2$Se$_3$, Bi$_2$Te$_3$ and Sb$_2$Te$_3$}
\label{sec:Te_transitions}
We are now in position to investigate the common features and differences of the topological phase transitions in the entire family of Bi/Sb Se/Te materials. We continue to use the ratio $c/a$ as a tuning parameter that separates the effects of the hydrostatic pressure from that of strain and stress, and allows a more direct analysis of the competing energy scales. Bi$_2$Te$_3$ and Sb$_2$Te$_3$ are known~\cite{Chen2009,Hsieh2009b} to have inverted bands in their ground state, and we confirmed this in our calculations. When we allow the systems to fully relax, we find that the ground states of Bi$_2$Te$_3$ and Sb$_2$Te$_3$ occur at $c/a$ = 7.18 and $c/a$ = 7.29 respectively~\cite{SI}. The optimized ground-state volumes differ by 0.4\% and 0.7\% from their respective experimental values~\cite{Nakajima1963,Anderson1974}.

Upon increasing $c/a$, we find phase transitions to trivial insulating states analogous to that in Bi$_2$Se$_3$ in all three materials, Sb$_2$Se$_3$, Bi$_2$Te$_3$ and Sb$_2$Te$_3$. While the QLs do not change substantially, the inter-QL distances increase significantly as $c/a$ increases: the changes in the inter-QL distances are two orders of magnitude greater than those of the bond lengths, see the detailed results in the Supplementary Material~\cite{SI}. From these we determined the critical values of $c/a$ as $\approx$ 7.46, 8.15, 7.86 for Sb$_2$Se$_3$, Bi$_2$Te$_3$ and Sb$_2$Te$_3$ respectively.


Similar to Bi$_2$Se$_3$, to elucidate the role of SOC and vdW interactions, we carried out the calculations where we turned them off for comparison. For example, in Sb$_2$Se$_3$ we compared the results for the cases of including (i) SOC and vdW, (ii) vdW but no SOC, (iii) SOC but not vdW, and (iv) neither SOC nor vdW, see the Supplementary Material~\cite{SI}. We find that the SOC has negligible effect on the lattice constants, and the optimized lattice constants for a given $c/a$ value for cases (i) and (ii) are almost identical, as are (iii) and (iv). The bond lengths differ by less than $\sim$0.03\%. The optimized inter-QL distance is only slightly more sensitive to the spin-orbit coupling: it is $\sim$0.5\% smaller with SOC  than when SOC is not included. This is consistent with the picture of spin-orbit as a local interaction. It is responsible for the band inversion, but does not affect the lattice constants significantly.

In contrast, the vdW correction changes the lattice constants: for a fixed $c/a$, both $c$ and $a$ decrease by $\sim$1.9\%, while the inter-QL distance decreases by $\sim$7.3\%, with the bond lengths decreasing by $\sim$0.8\%. Thus, the vdW correction has a significant effect on the structural parameters. We verified that vdW interaction only changes the electronic structure via its effect on the lattice: if we compute the electronic structure for a fixed set of lattice parameters with and without vdW interaction, the band gap and other electronic parameters remain essentially unchanged.

\begin{figure}
    \centering
    \includegraphics[width=1.0 \columnwidth,trim={2cm 0.5cm 2cm 1.5cm},clip]{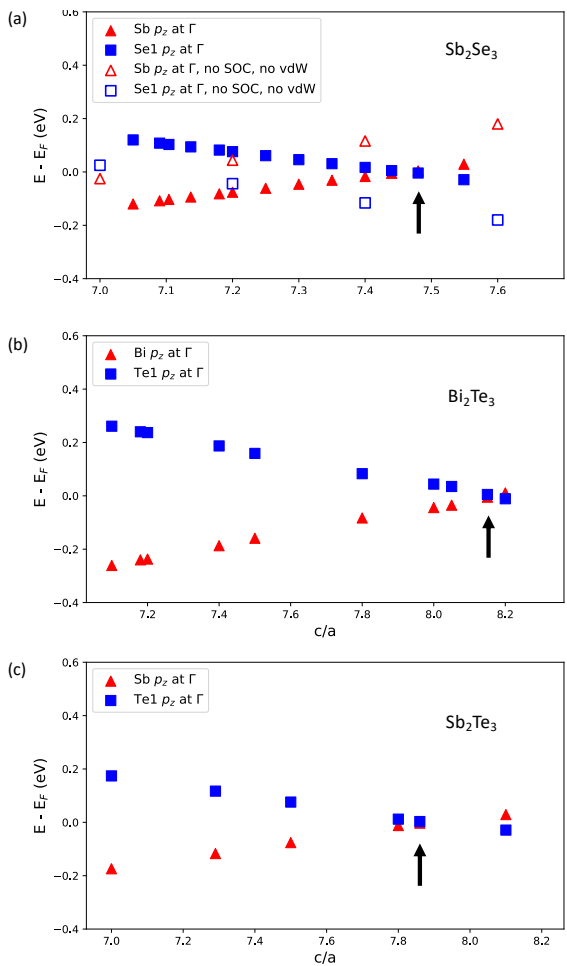}
    \caption{Dependence of the energies of the Bi/Sb $p_z$ and Te/Se $p_z$ states on $c/a$ in Sb$_2$Se$_3$, Bi$_2$Te$_3$ and Sb$_2$Te$_3$ at $\Gamma$. Red (blue) filled markers show the energies of the Bi/Sb $p_z$ (Te/Se $p_z$) levels for calculations including SOC and vdW interactions. Black arrows indicate the closing of the band gap and occurrence of the topological phase transition. (a) The red and blue unfilled markers show the energy levels of Sb$_2$Se$_3$ in calculations including neither SOC nor vdW. (b) Bi$_2$Te$_3$. (c) Sb$_2$Te$_3$. 
    }
    \label{fig:energy_levels_Te}
\end{figure}

We summarize these results in Figure~\ref{fig:energy_levels_Te}, which shows the energies of the Bi/Sb $p_z$ and Se/Te $p_z$ states at $\Gamma$, as a function of $c/a$. In analogy to Fig.~\ref{fig:transition_Bi2Se3}(c) we set the chemical potential in the middle of the gap in all cases. While in Sb$_2$Se$_3$, just as in its Bi counterpart, only the absolute energy of the band states originating from the outer Se layers change with $c/a$, in the Te compounds both Te and Bi/Sb $p_z$ level energies vary despite the QL layers and bond lengths remaining essentially unchanged~\cite{SI}.  This is likely due to the closer electronegativities and atomic sizes of Bi/Sb and Te, and leading to a more covalent (less ionic) bond between them. As a result, the shift in the energy of the electrons in the outer atomic sheet of the QL is reflected in the corresponding change of Bi/Sb levels. Plotting the energies relative to the mid-gap allows comparison across all four members of the family.

In analogy to Bi$_2$Se$_3$, we find that the energies of the states derived from the Se/Te and the Bi/Sb $p$-electrons in all three materials vary linearly with $c/a$. Bi$_2$Te$_3$ has the largest initial splitting between the Bi and Te $p_z$ levels at $\Gamma$ of $\sim$ 0.5 eV. While the energy and relative position of the levels depend on the inclusion of SOC and vdW, these only weakly influence the slope of the line. For example, in Sb$_2$Se$_3$, this slope is 0.294 $\pm$ 0.002 eV (0.344 $\pm$ 0.006 eV) with (without) including SOC+vdW, relatively close to the value for Bi$_2$Se$_3$. We find that the slopes of the Te $p_z$ levels in the Te compounds are smaller than those for the Se $p_z$ levels in the Se-based TIs: -0.243 $\pm$ 0.003 eV for Bi$_2$Te$_3$ and -0.190 $\pm$ 0.007 eV for Sb$_2$Te$_3$. This suggests that the evolution of the material towards the topological phase transition depends  more  on the nature of the elements in the outer atomic planes of the quintuple layers (Se and Te).

\section{Discussion}
\label{sec:discussion}
\begin{figure}
    \centering
    \includegraphics[width=\columnwidth,trim={0.75cm 4cm 1cm 6cm},clip]{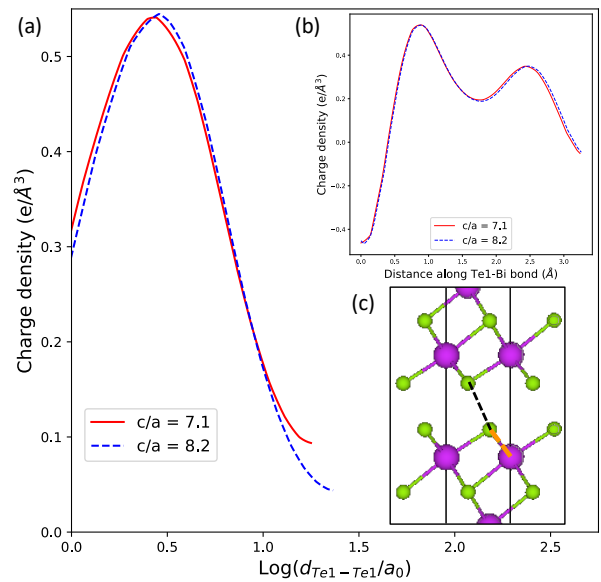}
    \caption{(a) Charge density of Bi$_2$Te$_3$ along line connecting Te atoms in neighboring QLs for $c/a$ = 7.1 (red, solid line) and $c/a$ = 8.2 (blue, dashed line), plotted up until half the atom spacing. (b) Inset shows charge densities along Te1-Bi bonds for $c/a$ = 7.1 (red, solid line) and $c/a$ = 8.2 (blue, dashed line). (c) Inset shows black dashed line connecting Te1 atoms between adjacent QLs ($d_{Te1-Te1}$), and orange dashed line along Te1-Bi bond.}
    \label{fig:charge_density}
\end{figure}

Our results show that the topological phase transitions, driven by tuning the ratio of the out-of-plane to in-plane lattice constants, $c/a$, exhibit common features in all of the chalcogenides Bi$_2$Se$_3$, Sb$_2$Se$_3$, Bi$_2$Te$_3$ and Sb$_2$Te$_3$. Structurally, as we increase this ratio, the lattice is elongated along the $c$-axis, with most of the change in the $c$-axis lattice constant due to the increase in the inter-QL distance. The compression that happens in the plane to the QLs is accompanied by small changes in the QL thickness, but the changes in the bond lengths within the QL are almost two orders of magnitude smaller than the corresponding variations in the inter-QL distances in all these compounds. The charge density along the line connecting Te1 atoms in adjacent QLs of Bi$_2$Te$_3$ exhibits small rearrangement for two different values of $c/a$ corresponding to topological and trivial states,  with faster drop-off for greater value of $c/a$, as shown in Fig.~\ref{fig:charge_density}(a).  Meanwhile the charge density along the Bi-Te1 bond within the QL remains unchanged, see Fig.~\ref{fig:charge_density}(b). Hence, structurally and electronically, the quintuple layers remain unchanged across the topological transition.

The topological transition is manifested in vanishing and reopening of the band gap, with the band inversion disappearing simultaneously. Electronic band structure calculations show that near the $\Gamma$ point the energy gap, as well as the Bi/Sb-derived and the Se/Te-derived $p_z$-states depend quasi-linearly on $c/a$ in the vicinity of the transition.  In the topological state, these Bi/Sb-derived (Se/Te-derived) sit at the top of the valence (bottom of the conduction) band, while their order is reversed in the trivial state. The band inversion is due to the spin-orbit coupling, and without it the Se/Te $p_z$-states are always lower in energy than the Bi/Sb $p_z$-states.

Remarkably, in all four materials we find that the the slope of that linear dependence does not depend strongly on the inclusion of spin-orbit coupling. This is consistent with the picture of the quintuple layers being unchanged as $c/a$ is varied, so that the SOC which originates mostly from Bi/Sb in the inner atomic layers is unaffected. However, our result contrasts with the scenario suggested in Ref.~\onlinecite{Liu2011}, where the strength of the SOC was argued to change with the inter-QL distance. We find no evidence of this.

We also find that the same slope is slightly affected by the inclusion of the van der Waals corrections, mostly via the change in the lattice parameters and the interlayer distance. 

The emergent picture is that the initial value of the band gap, at the ground state, is strongly influenced by the interplay of the van der Waals interactions between the quintuple layers and the spin-orbit-induced band inversion. However, the evolution from that state towards the topological transition is influenced primarily by the inter-QL physics, which is determined mostly by the Coulomb interaction (contained in the density functional) with corrections due to the van der Waals interactions. 

Consider, for example, the slope of the energy of the Se-derived states in Sb$_2$Se$_3$ obtained using different DFT-based methods, shown in Table~\ref{slopes_vs_c_by_a_Sb2Se3}, based on the data in the Supplemental Materials~\cite{SI}. We find that the inclusion of SOC does not change the energy evolution of these states, while inclusion of van der Waals forces does, albeit weakly. The relative decrease in the slope upon including vdW supports our view of the dominant role of the inter-QL physics. PBE-GGA often underbinds, and hence, upon optimization, the Coulomb repulsion due to the overlapping electron clouds of the nearest neighbors in different QLs and the gradient terms cause the inter-QL spacing to increase, decreasing the band gap, and lowering the energy of the Se $p_z$ level. In contrast, the inclusion of the attractive van der Waals interaction reduces $d_{QL}$ when compared to that obtained by PBE-GGA, and increases the band gap (thereby increasing the energy of the Se $p_z$ level). Therefore the slope of the dependence of the energy of Se $p_z$ levels on $c/a$ is smaller with the vdW corrections. Thus, it is the inter-QL coupling which influences how the energies of the Bi/Sb- and Se/Te-derived $p$-states at the $\Gamma$ point vary with $c/a$. Finally, we also note that the minor differences between the slopes in Table~\ref{slopes_vs_c_by_a_Sb2Se3} with and without SOC are due to the small differences between the optimized inter-QL distances in each case.

\begin{table}
\begin{tabular}{| l | c |}
\hline
Method & 
Slope II (eV) \\
\hline
PBE-GGA + vdW + SOC &   0.294$\pm$0.002 \\
PBE-GGA + vdW, no SOC & 0.290$\pm$0.004 \\ 
PBE-GGA + SOC, no vdW  & 0.346$\pm$0.008 \\ 
PBE-GGA, no vdW, no SOC & 0.344$\pm$ 0.006 \\ 
\hline
\end{tabular}
\caption{\label{slopes_vs_c_by_a_Sb2Se3} Slopes of Se $p_z$ levels in Sb$_2$Se$_3$ as a function of $c/a$ for calculations incorporating (i) vdW + SOC, (ii) vdW but no SOC, (iii) SOC but no vdW, and (iv) neither vdW nor SOC. 
}
\end{table}

We find that the inter-QL distances $d_{QL}$ also vary linearly with $c/a$ (see also Ref.~\onlinecite{Reid2021}), and find that the corresponding slopes
for the two Se compounds are almost identical, as are those of the two Te compounds (see second column in Table~\ref{distances_vs_c_by_a}, and the full data in the Supplemental Information~\cite{SI}). This shows that the changes in $d_{QL}$ are determined by the type of atom in the outermost planes of the quintuple layers, Te or Se.

\begin{table}
\begin{tabular}{| c | c | c | c |}
\hline
Material & 
{Slope interlayer} & \multicolumn{2}{|c|}{Critical value} \\
\hline
 & distance $d_{QL}$  & $d_{y{\text -}y}$ ({\AA}) & ($d_{y{\text -}y} - r_{y}^{ion}$) ({\AA}) \\
\hline
Bi$_2$Se$_3$ & 0.66  & 3.7 & 2.7 \\
Sb$_2$Se$_3$ & 0.67  & 3.7 & 2.7 \\
Bi$_2$Te$_3$ & 0.58  & 4.1 & 2.9 \\ 
Sb$_2$Te$_3$ & 0.57  & 4.0 & 2.8 \\ 
\hline
\end{tabular}
\caption{\label{distances_vs_c_by_a} Rate of change of inter-QL distances 
as a function of $c/a$. Last two columns show the critical value of atom spacings between QLs $d_{y{\text -}y}$ ($y$ = Te/Se), and critical atom spacings between QLs minus the corresponding ionic radii of Te/Se. }
\end{table}

If we focus on the distances between atoms in adjacent QLs, $d_{y{\text -}y}$ ($y$ = Te/Se, see Fig.~\ref{fig:charge_density}(c) for illustration), at the value of $c/a$ where the topological transition occurs in each material, we find that they are identical for the two Se-based compounds, and also identical, albeit at a different value, for the two Te-based materials, see Table~\ref{distances_vs_c_by_a}. Change of Bi for Sb does not affect this distance. Moreover, if we account for the difference in the ionic radii between Se and Te, we find that the topological transition in all four compounds occurs at approximately the same ``critical'' separation between the QLs, $\sim$ 2.8 {\AA}. This gives predictive power to our analysis, as measurements of the changes in the lattice constants in a given experimental setup can be extrapolated to the ``universal'' critical value to predict the parameters for the topological transition.

\begin{table}
\begin{tabular}{| c |c | c |c|}
\hline
Material & \multicolumn{3}{|c|}{Slope with respect to $c/a$ of} \\
\hline
 & Y1 $p_z$ at $\Gamma$ (eV) & $d_{QL}^\prime/d_{QL}$ & Band gap $E_g$ (eV) \\
\hline
Bi$_2$Se$_3$ & 0.31 & 0.259  & -0.58 $\pm$ 0.01 \\
Sb$_2$Se$_3$ & 0.29 & 0.245 & -0.46 $\pm$ 0.01 \\
Bi$_2$Te$_3$ & 0.24 & 0.210 & -0.38 $\pm$ 0.01 \\ 
Sb$_2$Te$_3$ & 0.19 & 0.195 & -0.39 $\pm$ 0.01 \\ 
\hline
\end{tabular}
\caption{\label{energy_slopes_vs_c_by_a} The slope of linear change in the chalcogenide $p_z$ energy levels (Y1 = Se1/Te1) and band gaps $E_g$ as a function of $c/a$ correlates with the relative variation of the interlayer distance.}
\end{table}

We find that the energetics of the $p_z$ levels supports the crucial role of the interlayer coupling. Among the four materials, Bi$_2$Se$_3$ displays the largest variation in the $p_z$ level energies with respect to changing the $c/a$ ratio, followed by Sb$_2$Se$_3$ and Bi$_2$Te$_3$, while Sb$_2$Te$_3$ displays the least variation, see the second column in Table~\ref{energy_slopes_vs_c_by_a}. We conjecture that the large electronegativity difference between Bi and Se (0.53), which implies significant electron density closer to the Se atom in the Bi-Se bond, makes the Coulomb repulsion between Se atoms in different QLs stronger. In contrast, Sb and Te have close electronegativity (difference 0.05), which results in a more covalent distribution of the electrons between the two, smaller electron densities near the outermost Te, and hence weaker sensitivity to changing the inter-QL distance. Fitting the band gaps linearly yields slopes for the Te compounds which are very close, whereas the corresponding slopes for the Se compounds have a larger difference, see the third column in Table~\ref{energy_slopes_vs_c_by_a}. The band gaps in the four materials are close in value to the separation between the Bi/Sb and Te/Se $p_z$ level energies at $\Gamma$. However, the `camelback' structure in the valence band causes the band gap to be smaller than this separation. Small changes in the camelback structure causes the slope of the band gap to differ slightly from the slopes of the $p_z$ levels in Table~\ref{energy_slopes_vs_c_by_a}.

Both the variation of the energy gap and the change in the $p_z$ level energy correlate with the normalized rate of change of the inter-QL distance, $d_{QL}^\prime/d_{QL}$ as shown Table~\ref{energy_slopes_vs_c_by_a}. The ground-state inter-QL distance is the smallest in Bi$_2$Se$_3$, with Sb$_2$Se$_3$, Bi$_2$Te$_3$ and Sb$_2$Te$_3$ following in order of increasing $d_{QL}$. The materials with the greater relative change in the interlayer distance exhibit stronger dependence of the energy of the band states near the $\Gamma$ point. This strongly supports the view that the inter-QL physics is responsible for the evolution of these energy levels.

\section{Conclusions}

While our results are in general agreement with interpreting the transition from topological to a trivial insulator as a consequence of the competition between level splitting due to hybridization and spin-orbit coupling that promotes band inversion, the underlying mechanism of inter-QL interactions is different from that suggested previously.  Refs.~\onlinecite{Liu2010, Zhang2009} considered the atomic $p$ orbitals of Bi and Se in a single QL, and suggested the steps towards band inversion that were largely based on intra-QL physics. They first included crystal field splitting,  hybridization, and level repulsion between Bi and Se, followed by inclusion of SOC. Our main conclusion here is that the QLs remain largely unchanged, and therefore SOC and Bi-Se hybridization strength do not vary almost at all within the considered range of the lattice constants. In that language, the ligand-induced splitting of the $p$-electron manifold of Se/Te is the driving force behind the topological phase transition, with the important caveat that the changes in this splitting are almost entirely due to inter-QL interaction.  While the pressure effect on the splitting of the energy levels obviously has been identified before~\cite{Liu2011,Li2014} we are not aware of any work making unambiguous connection of the topological transition with the inter-QL physics.


\begin{figure}
    \centering
    \includegraphics[width=\columnwidth]{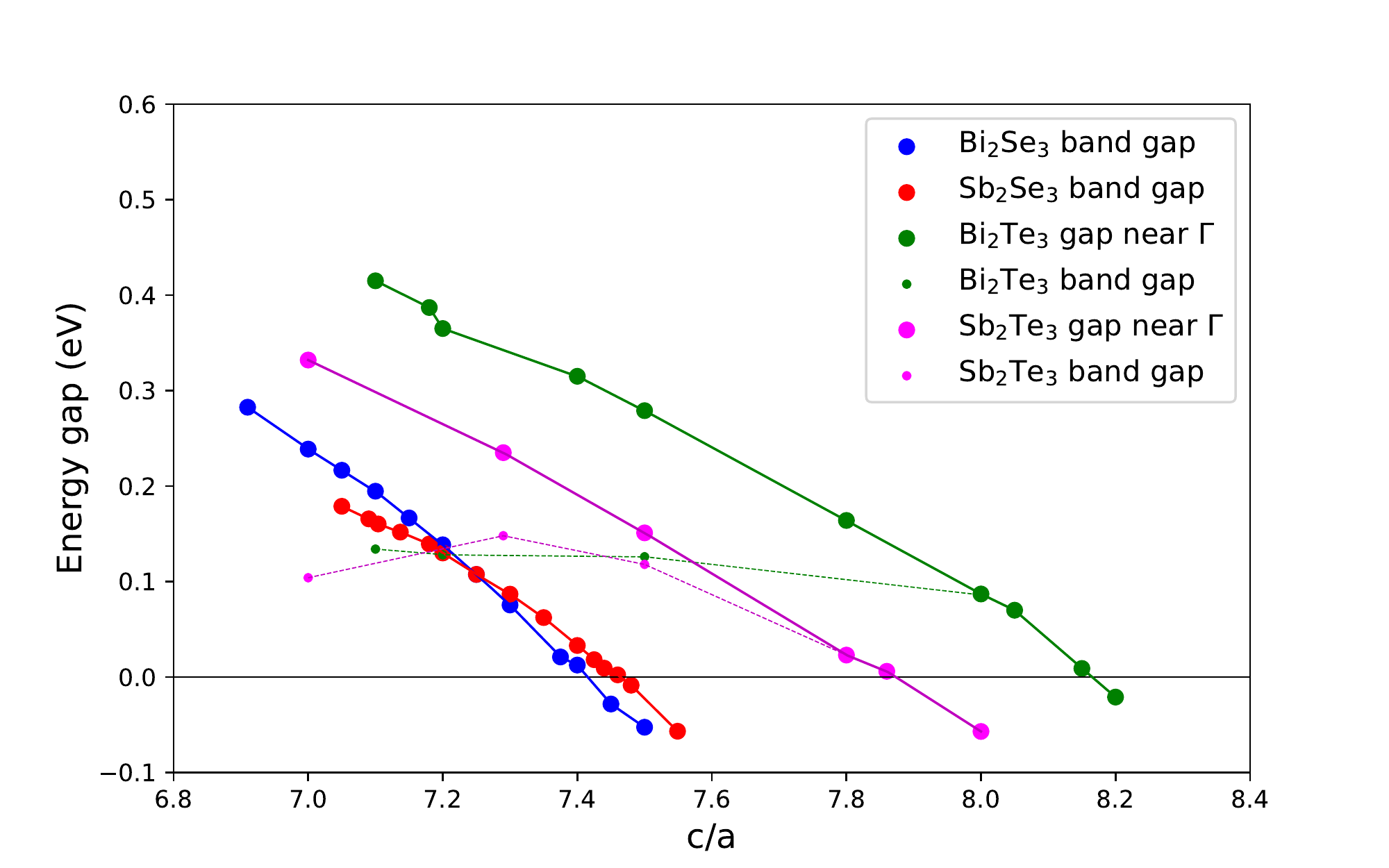}
    \caption{Bulk band gaps and gaps near $\Gamma$ of Bi$_2$Se$_3$, Sb$_2$Se$_3$, Bi$_2$Te$_3$ and Sb$_2$Te$_3$ as a function of $c/a$.}
    \label{fig:band_gap_vs_c_by_a}
\end{figure}

Our results allow us to make conjectures about 
the topological phase transitions in the alloys of the materials we studied. The original interest in alloying arose because, with Sb$_2$Se$_3$ predicted to be a trivial insulator, the alloy (Bi$_{1-x}$Sb$_x$)$_2$Se$_3$ was expected~\cite{Liu2013} to undergo a topological phase transition (TPT) at a critical value of $x$. Given the current results showing that this end material is a topological insulator, the question of whether the transition in the alloys exists, and, if yes, what its nature is, remains open, and will be addressed by us in detail in future work.

In Fig.~\ref{fig:band_gap_vs_c_by_a}, we show the variation of the gaps near $\Gamma$ as a function of $c/a$. Note that, while in the Se compounds the minimal gap is always near the $\Gamma$ point, we find that, away from the critical value of $c/a$, the gap in the Te compounds is not at the center of the Brillouin Zone. Hence we plot the gap in the density of states and the gap at the $\Gamma$ point separately, as only the latter is relevant for the topological transition. 
We can think of alloying as both producing the chemical pressure and introducing disorder in the on-site energies of the $p_z$ levels. The critical $c/a$ values of Bi$_2$Se$_3$ and Sb$_2$Se$_3$ are very close, and hence it is not likely that (Bi-Sb) alloying in Se-based material at ambient pressure yields a topological phase transition. It is possible, of course, that the difference in the SOC strength on Bi and Sb conspires with the optimized lattice constants at a given $x$ to remove the band inversion, but this would be accidental fine-tuning of parameters in the bulk. Ref.~\onlinecite{Liu2013} reported that in (Bi$_{1-x}$Sb$_x$)$_2$Se$_3$ semimetallic states appear over a range of concentrations between $x = 0.78$ and $x = 0.83$, rather than at the single transition point, as would be expected. Hence it is important to revisit this issue.

Experiments~\cite{Satake2018} have shown the existence of the Dirac surface states in (Bi$_{1-x}$Sb$_x$)$_2$Se$_3$ grown on Bi$_2$Se$_3$ up to at least $x$=0.7. Note that we performed a bulk calculation with full relaxation of the volume for each value of $c/a$. In thin films grown on a substrate the in-plane lattice constant is fixed, and hence the critical inter-QL distance will differ from that in the bulk. We also leave this for future studies, and are currently carrying out calculations to investigate the topological nature of the alloys under these conditions.   

Returning to the bulk systems, the topological transitions for Bi$_2$Te$_3$ and Sb$_2$Te$_3$ happen at quite different $c/a$ values, and an alloy of the two could thus display a TPT as the concentrations of Bi and Sb atoms are varied. For the same reason, we suggest that it might be possible to observe a TPT in Bi$_2$(Te$_y$Se$_{1-y}$)$_3$ by varying the concentration $y$. In fact, additional work is needed to determine whether such alloying yields a different concentration of Se and Te in the middle and outer atomic sheets of the quintuple layers, and whether it induces reconstruction of the outer atomic layers, which would, according to our results, have a strong effect on the topological properties.

In summary, we investigated the topological transition in stoichiometric tetradymite topological insulators using the ratio of the lattice constants as a tuning parameter. Our main conclusion is that in all the materials studied this transition is driven by the inter-QL Coulomb and van der Waals interactions between the outer atoms in the quintuple layers, and therefore the inter-QL distance is the key tuning parameter that controls this transition. We identified van der Waals forces and the nature of the intra-QL bonding that determines the charge distribution between the layers as main factors in controlling the approach to such transitions. Our results suggest pathways towards realization of the topological phase transitions in bulk materials and thin films.

\acknowledgments{This research was supported by NSF via Grant No. DMR-1410741 (K. S. and I. V.) and by the U.S. Department of Energy under EPSCoR Grant No. DE-SC0012432 with additional support from the Louisiana Board of Regents (W.~A.~S. and K. S.). I. V. is grateful to the Kavli Institute for Theoretical Physics for its hospitality during early stages of this work, supported in part by the National Science Foundation under Grant No. NSF PHY-1748958. Portions of this research were conducted with high performance computing resources provided by Louisiana State University (\url{http://www.hpc.lsu.edu}).}


%

\end{document}